\documentclass[nofootinbib,aps]{revtex4}%
\usepackage{amsmath}
\usepackage{amsfonts}
\usepackage{amssymb}
\usepackage{graphicx}%
\begin{document}
\title{How Spacetime Foam modifies the brick wall.}
\author{Remo Garattini}
\email{Garattini@mi.infn.it - Remo.Garattini@unibg.it}
\affiliation{Viale Marconi 5, 24044 Dalmine (Bergamo) ITALY.}

\begin{abstract}
We re-examine the brick-wall model in the context of spacetime foam. In
particular we consider a foam composed by wormholes of different sizes filling
the black hole horizon. The contribution of such wormholes is computed via a
scale invariant distribution. We obtain that the brick wall divergence appears
to be logarithmic when the cutoff is sent to zero.

\end{abstract}
\maketitle

\section{Introduction}

\label{p1}Gerard `t Hooft\cite{tHooft}, in 1985 considered the statistical
thermodynamics of quantum fields in the Hartle-Hawking state (i.e. having the
Hawking temperature $T_{H}$ at large radii) propagating on a fixed
Schwarzschild background of mass $M$. To control divergences, `t Hooft
introduced a \textquotedblleft\textit{brick wall}\textquotedblright\ with
radius a little larger than the gravitational radius $2MG$. He found, in
addition to the expected volume-dependent thermodynamical quantities
describing hot fields in a nearly flat space, additional contributions
proportional to the area. These contributions are, however, also proportional
to $\alpha^{-2}$, where $\alpha$ is the proper distance from the horizon, and
thus diverge in the limit $\alpha\rightarrow0$. For a specific choice of
$\alpha$, he recovered the Bekenstein-Hawking formula%
\begin{equation}
S_{BH}=\frac{1}{4}A/l_{P}^{2}. \label{eqn:S_BH}%
\end{equation}
The prescription for assigning a \textquotedblleft Bekenstein-Hawking
entropy\textquotedblright\ $S_{BH}$ to a black hole of surface area $A$ was
first inferred in the mid-1970s from the formal similarities between black
hole dynamics and thermodynamics\cite{Bekenstein}, combined with Hawking's
discovery\cite{Hawking} that black hole radiate thermally with a
characteristic (Hawking) temperature%
\begin{equation}
T_{H}=\hbar\kappa_{0}/2\pi, \label{eqn:T_H}%
\end{equation}
where $\kappa_{0}$ is the surface gravity. Since then, many attempts to
renormalize or eliminate the \textquotedblleft\textit{brick wall}%
\textquotedblright\ have been done. In a series of paper, it has been
suggested\cite{SusUgl,BarEmp,EWin} that this divergence could be absorbed in a
renormalization of Newton's constant, while other authors approached the
problem of the divergent brick wall using Pauli-Villars
regularization\cite{DLM,FS,KKSY}. In the Pauli-Villars covariant
regularization method, one introduces bosonic and fermionic regulator fields
to regulate the divergences. What happens is that the free energy of the
anti-commuting regulator fields comes with a minus sign with respect to the
commuting fields. This leads to a cancellation of the ultraviolet divergence
when the 't Hooft brick wall is removed. Recently a proposal coming by the
modification of the Heisenberg uncertainty relations has been taken under
consideration\cite{Xiang Li,RenQinChun}. The modified inequality takes the
form\footnote{Consequences of this modification have been discussed in
Refs.\cite{KMM,Garay,Ahluwalia,ACS,Rama,CMOT}}%
\begin{equation}
\Delta x\Delta p\geq\hbar+\frac{\lambda_{p}^{2}}{\hbar}\left(  \Delta
p\right)  ^{2},
\end{equation}
where $\hbar$ is the Planck constant and $\lambda_{p}$ is the Planck length.
The interesting point regards exactly the modified number of quantum states,
which is changed into%
\begin{equation}
\frac{d^{3}xd^{3}p}{\left(  2\pi\hbar\right)  ^{3}\left(  1+\lambda
p^{2}\right)  ^{3}}. \label{eqn:states}%
\end{equation}
When $\lambda=0$, the formula reduces to the ordinary counting of quantum
states. If Eq.$\left(  \ref{eqn:states}\right)  $ is used for computing the
entropy, the brick wall can be removed. Another interesting recent proposal
comes from non-commutative geometry which introduces a natural thickness of
the horizon replacing the 't Hooft's brick wall\cite{BaiYan}. In this paper we
wish to repeat the brick wall computation in the context of spacetime foam. It
was J. A. Wheeler who first conjectured that spacetime could be subjected to
topology fluctuation at the Planck scale\cite{Wheeler}. These fluctuations
appearing at this scale form the \textquotedblleft\textit{spacetime
foam}\textquotedblright. An interesting calculation scheme in this context
comes by L. Crane and L. Smolin\cite{CraneSmolin}. They show that in a foamy
spacetime, general relativity can be renormalized when a density of virtual
black holes coupled to $N$ fermion fields in a $1/N$ expansion is taken under
consideration. The idea they propose is that the high-energy behavior of
perturbation theory can be modified by the presence of nonperturbative
structure in the vacuum at arbitrary small length scales. In their work, they
used a distribution of virtual black holes $\rho\left(  w\right)  $ suggested
by the uncertainty principle with the assumption that the perturbations of
matter and gravitational fields contributing as virtual states in perturbation
theory vanish on and inside any apparent horizons present in the background
spacetime. Thus they proposed that in the spacelike slice $\Sigma$ of the
background geometry, the distribution $\rho\left(  w\right)  dw$ per unit
available volume (including also the case of an interior of a black hole of
larger size) with masses between $w$ and $w+dw$ was of the form%
\begin{equation}
\rho\left(  w\right)  =\left\{
\begin{array}
[c]{cc}%
0 & w>l_{p}\\
\frac{C}{w^{4}}\left(  \frac{l_{p}}{w}\right)  ^{q} & w<l_{p},
\end{array}
\right.  \label{eqn:density}%
\end{equation}
where $C$ is a dimensionless constant and $q$ is a parameter such that
$q\gtreqless0$. At this point a natural question arises: how can we relate the
Crane-Smolin picture of foam with distribution $\rho\left(  w\right)  $ with
the brick wall? Motivated by a recent proposal of a model of spacetime foam
based on a superposition of wormholes\cite{RIJMPD}, we explore the possibility
that such a non trivial configuration can affect the behavior of a quantum
field near the black hole horizon. The main idea is to substitute the
superposition of commuting and anti-commuting fields of
Refs.\cite{DLM,FS,KKSY} renormalizing the entropy with the gravitational field
itself and to use the distribution $\rho\left(  b\right)  $ in $\left(
\ref{eqn:density}\right)  $ to deal with the modified number of quantum states
of Eq.$\left(  \ref{eqn:states}\right)  $. This combination could change the
divergent behavior of the entropy when the brick wall is removed. The rest of
the paper is structured as follows, in section \ref{p2} we recall the
fundamental points that lead to a brick wall, in section \ref{p3} we apply the
foam model to the brick wall model. We summarize and conclude in section
\ref{p4}. Units in which $\hbar=c=k=1$ are used throughout the paper.

\section{Brick wall model}

\label{p2}We wish to study the thermodynamics of hot quantum fields in a
background geometry of the form\cite{Visser}%
\begin{equation}
ds^{2}=-\exp\left(  -2\Lambda\left(  r\right)  \right)  \left(  1-\frac
{b\left(  r\right)  }{r}\right)  dt^{2}+\frac{dr^{2}}{\left(  1-\frac{b\left(
r\right)  }{r}\right)  }+r^{2}d\Omega^{2}. \label{e31}%
\end{equation}
Usually this form is considered for the description of wormholes. However, it
is quite general to include as special cases the Schwarzschild,
Reissner-Nordstr{\"{o}}m and de Sitter geometries, or any combination of
these. The function $b\left(  r\right)  $ will be referred to as the
\textquotedblleft shape function\textquotedblright. The shape function may be
thought of as specifying the shape of the spatial slices. On the other hand,
$\Lambda\left(  r\right)  $ will be referred to as the \textquotedblleft
redshift function\textquotedblright\ that describes how far the total
gravitational redshift deviates from that implied by the shape function.
Without loss of generality we can fix the value of $\Lambda\left(  r\right)  $
at infinity such that $\Lambda\left(  \infty\right)  =0$. If the equation
$b\left(  r_{w}\right)  =r_{w}$ is satisfied for some values of $r$, then we
say that the points $r_{w}$ are horizons for the metric $\left(
\ref{e31}\right)  $. For the outermost horizon one has $\forall r>r_{w}$ that
$b\left(  r\right)  <r$. Consequently $b^{\prime}\left(  r_{w}\right)  \leq1$.
We will fix our attention to the $b^{\prime}\left(  r_{w}\right)  <1$ case
only. The anomalous case $b^{\prime}\left(  r_{w}\right)  =1$ can be thought
as describing extreme black holes where an inner and outer horizons are
merged. For a spherically symmetric system the surface gravity is computed
via
\begin{equation}
\kappa_{w}=\lim_{r\rightarrow r_{w}}\left\{  \frac{1}{2}\frac{\partial
_{r}g_{tt}}{\sqrt{g_{tt}g_{rr}}}\right\}
\end{equation}
and for the metric $\left(  \ref{e31}\right)  $, we get
\begin{equation}
\kappa_{w}=\lim_{r\rightarrow r_{w}}\frac{1}{2}\left\{  \exp\left(
-\Lambda\left(  r\right)  \right)  \left[  -2\Lambda^{\prime}\left(  r\right)
\left(  1-\frac{b\left(  r\right)  }{r}\right)  +\frac{b\left(  r\right)
}{r^{2}}-\frac{b^{\prime}\left(  r\right)  }{r}\right]  \right\}  .
\end{equation}
By assuming that $\Lambda\left(  r_{w}\right)  $ and $\Lambda^{\prime}\left(
r_{w}\right)  $ are both finite we obtain that
\begin{equation}
\kappa_{w}=\frac{1}{2r_{w}}\exp\left(  -\Lambda\left(  r_{w}\right)  \right)
\left[  1-b^{\prime}\left(  r_{w}\right)  \right]  , \label{e32}%
\end{equation}
where, in the proximity of the throat we have approximated $1-b\left(
r\right)  /r$ with%
\begin{equation}
1-\frac{b\left(  r\right)  }{r}=\frac{r-r_{w}}{r_{w}}\text{ }\left[
1-b^{\prime}\left(  r_{w}\right)  \right]  . \label{e32a}%
\end{equation}
Now that the geometrical framework has been set up, we begin with a real
massless scalar field described by the action\footnote{see also
Ref.\cite{MukohyamaIsrael} for a derivation of the brick wall in the Boulware
state.}
\begin{equation}
I=-\frac{1}{2}\int d^{4}x\sqrt{-g}\left[  g^{\mu\nu}\partial_{\mu}\phi
\partial_{\nu}\phi\right]
\end{equation}
in the background geometry of Eq.$\left(  \ref{e31}\right)  $ whose
Euler-Lagrange equations are%
\begin{equation}
\frac{1}{\sqrt{-g}}\partial_{\mu}\left(  \sqrt{-g}g^{\mu\nu}\partial_{\nu
}\right)  \phi=0.
\end{equation}
If $\phi$ has the separable form
\begin{equation}
\phi\left(  t,r,\theta,\varphi\right)  =\exp\left(  -i\omega t\right)
Y_{lm}(\theta,\varphi)f\left(  r\right)  ,
\end{equation}
then the equation for $f\left(  r\right)  $ reads%
\begin{equation}
\frac{\exp\left(  \Lambda\left(  r\right)  \right)  }{r^{2}}\partial
_{r}\left(  r^{2}\exp\left(  -\Lambda\left(  r\right)  \right)  \left(
1-\frac{b\left(  r\right)  }{r}\right)  \partial_{r}f_{nl}\right)  -\left[
\frac{l(l+1)}{r^{2}}\right]  f_{nl}+\omega_{nl}^{2}\frac{\exp\left(
2\Lambda\left(  r\right)  \right)  }{1-\frac{b\left(  r\right)  }{r}}f_{nl}=0,
\label{e33}%
\end{equation}
where $Y_{lm}(\theta,\varphi)$ is the usual spherical harmonic function. In
order to make our system finite let us suppose that two mirror-like boundaries
are placed at $r=r_{1}$ and $r=R$ with $R\ggg r_{1},$ $r_{1}>r_{w}$ and
consider Dirichlet boundary conditions $f_{nl}(r_{1})=f_{nl}(R)=0$. We also
assume the set of real functions $\{f_{nl}(r)\}$ ($n=1,2,\cdots$), defined by
Eq.$\left(  \ref{e33}\right)  $, be complete with respect to the space of
$L_{2}$-functions on the interval $r_{1}\leq r\leq R$ for each $l$. The
positive constant $\omega_{nl}$ is defined as the corresponding eigenvalue. In
order to use the WKB approximation, we define an r-dependent radial wave
number $k(r,l,\omega_{nl})$
\begin{equation}
k^{2}(r,l,\omega_{nl})\equiv\frac{1}{\left(  1-\frac{b\left(  r\right)  }%
{r}\right)  }\left[  \exp\left(  2\Lambda\left(  r\right)  \right)
\frac{\omega_{nl}^{2}}{\left(  1-\frac{b\left(  r\right)  }{r}\right)  }%
-\frac{l(l+1)}{r^{2}}\right]  .
\end{equation}
The number of modes with frequency less than $\omega$ is given approximately
by
\begin{equation}
\tilde{g}(\omega)=\int\nu(l{,\omega})(2l+1)dl,
\end{equation}
where $\nu(l,\omega)$ is the number of nodes in the mode with $(l,\omega)$:
\begin{equation}
\nu(l,\omega)=\frac{1}{\pi}\int_{r_{1}}^{R}\sqrt{k^{2}(r,l,\omega)}dr.
\end{equation}
Here it is understood that the integration with respect to $r$ and $l$ is
taken over those values which satisfy $r_{1}\leq r\leq R$ and $k^{2}%
(r,l,\omega)\geq0$. The free energy is given approximately by
\begin{equation}
F\simeq\frac{1}{\beta_{\infty}}\int_{0}^{\infty}\ln\left(  1-e^{-\beta
_{\infty}\omega}\right)  \frac{d\tilde{g}(\omega)}{d\omega}d\omega=\int
_{r_{1}}^{R}\tilde{F}(r)4\pi r^{2}dr,
\end{equation}
where the `free energy density' $\tilde{F}(r)$ is defined by
\begin{equation}
\tilde{F}(r)\equiv\frac{\exp\left(  -\Lambda\left(  r\right)  \right)  }%
{\beta(r)}\int_{0}^{\infty}\ln\left(  1-e^{-\beta(r)p}\right)  \frac{4\pi
p^{2}dp}{(2\pi)^{3}}.
\end{equation}
Here the \textquotedblleft local inverse temperature\textquotedblright%
\ $\beta(r)$ is defined by the Tolman's law\cite{Tolman}
\begin{equation}
\beta(r)=\exp\left(  -\Lambda\left(  r\right)  \right)  \sqrt{1-\frac{b\left(
r\right)  }{r}}\beta_{\infty}.
\end{equation}
The total entropy is given by the integral
\begin{equation}
S=4\pi\int_{r_{1}}^{R}\frac{s(r)}{\sqrt{1-\frac{b\left(  r\right)  }{r}}}%
r^{2}dr=\frac{16\pi^{3}}{90\beta_{\infty}^{3}}\int_{r_{1}}^{R}\frac
{\exp\left(  2\Lambda\left(  r\right)  \right)  }{\left(  1-\frac{b\left(
r\right)  }{r}\right)  ^{2}}r^{2}dr=S_{r_{1}}+S_{R},
\end{equation}
where%
\begin{equation}
S_{r_{1}}=\frac{16\pi^{3}}{90\beta_{\infty}^{3}}\int_{r_{1}}^{r_{1}%
+\varepsilon}\frac{\exp\left(  2\Lambda\left(  r\right)  \right)  }{\left(
1-\frac{b\left(  r\right)  }{r}\right)  ^{2}}r^{2}dr \label{e33a}%
\end{equation}
and%
\begin{equation}
S_{R}=\frac{16\pi^{3}}{90\beta_{\infty}^{3}}\int_{r_{1}+\varepsilon}^{R}%
\frac{\exp\left(  2\Lambda\left(  r\right)  \right)  }{\left(  1-\frac
{b\left(  r\right)  }{r}\right)  ^{2}}r^{2}dr \label{e33b}%
\end{equation}
with $R\gg r_{1}$ and $r_{1}\gg\varepsilon\gg r_{1}-r_{w}$. For large $R$,
$S_{R}$ is dominated by\footnote{Recall that $b\left(  r\right)  \leq r$ and
$\Lambda\left(  \infty\right)  =0$.}%
\begin{equation}
S\sim\frac{16\pi^{3}}{90\beta_{\infty}^{3}}\frac{R^{3}}{3},
\end{equation}
representing the entropy of a homogeneous quantum gas in flat space at a
uniform temperature $T_{\infty}$. However, the brick wall divergence is in the
integral of Eq.$\left(  \ref{e33a}\right)  $. If we set $r_{1}=r_{w}+h$, then
we are led to consider the following integral%
\begin{equation}
S_{brick}=\frac{16\pi^{3}}{90\beta_{\infty}^{3}}\int_{r_{w}+h}^{r_{w}%
+h+\varepsilon}\frac{\exp\left(  2\Lambda\left(  r\right)  \right)  }{\left(
1-\frac{b\left(  r\right)  }{r}\right)  ^{2}}r^{2}dr=S\left(  \varepsilon
,h,r_{w}\right)  ,
\end{equation}
where%
\begin{equation}
S\left(  \varepsilon,h,r_{w}\right)  =\frac{16\pi^{3}}{90\beta_{\infty}^{3}%
}\frac{\exp\left(  2\Lambda\left(  r_{w}\right)  \right)  r_{w}^{2}}{\left(
1-b^{\prime}\left(  r_{w}\right)  \right)  ^{2}}\int_{r_{w}+h}^{r_{w}%
+h+\varepsilon}\frac{r^{2}dr}{\left(  r-r_{w}\right)  ^{2}}=\frac{16\pi^{3}%
}{90\beta_{\infty}^{3}}\frac{1}{4\kappa_{w}^{2}}\int_{r_{w}+h}^{r_{w}%
+h+\varepsilon}\frac{r^{2}}{\left(  r-r_{w}\right)  ^{2}}dr.
\end{equation}
By defining%
\begin{equation}
g\left(  \varepsilon,h,r_{w}\right)  =\int_{r_{w}+h}^{r_{w}+h+\varepsilon
}\frac{r^{2}}{\left(  r-r_{w}\right)  ^{2}}dr \label{e34}%
\end{equation}
and with the help of Eq.$\left(  \ref{e32}\right)  $, we get%
\begin{equation}
S\left(  \varepsilon,h,r_{w}\right)  =\frac{16\pi^{3}}{90}\frac{g\left(
\varepsilon,h,r_{w}\right)  }{4\beta_{\infty,r_{w}}^{3}\kappa_{w}^{2}}.
\label{e34a}%
\end{equation}
The \textit{brick wall} is obtained by keeping only the leading divergence in
Eq.$\left(  \ref{e34}\right)  $ and introducing the proper distance from the
throat%
\begin{equation}
\alpha=\int_{r_{w}}^{r_{w}+h}\frac{dr}{\sqrt{1-\frac{b\left(  r\right)  }{r}}%
}=\frac{2\sqrt{h}}{\sqrt{\frac{1-b^{\prime}\left(  r_{w}\right)  }{r_{w}}}}.
\end{equation}
Thus Eq.$\left(  \ref{e34a}\right)  $ becomes%
\[
S\left(  \varepsilon,h,r_{w}\right)  =\frac{16\pi^{3}}{90\beta_{\infty}^{3}%
}\frac{r_{w}^{2}}{4\kappa_{w}^{3}}\frac{\exp\left(  -\Lambda\left(
r_{w}\right)  \right)  }{\alpha^{2}}=\frac{16\pi^{3}}{90}\left(
\frac{T_{\infty}}{\kappa_{w}/2\pi}\right)  ^{3}\frac{r_{w}^{2}}{\left(
2\pi\right)  ^{3}}\frac{\exp\left(  -\Lambda\left(  r_{w}\right)  \right)
}{2\alpha^{2}}%
\]%
\begin{equation}
=\frac{A}{90}\left(  \frac{T_{\infty}}{\kappa_{w}/2\pi}\right)  ^{3}\frac
{\exp\left(  -\Lambda\left(  r_{w}\right)  \right)  }{4\pi\alpha^{2}}.
\end{equation}
The area law is recovered if we make the following identifications%
\begin{equation}
T_{\infty}=\frac{\kappa_{w}}{2\pi} \label{e35}%
\end{equation}
and%
\begin{equation}
\frac{\exp\left(  -\Lambda\left(  r_{w}\right)  \right)  }{90\pi\alpha^{2}%
}=\frac{1}{l_{p}^{2}}. \label{e36}%
\end{equation}

\begin{description}
\item[\textbf{Remark 1}] In this paper the redshift function is considered
practically as a constant. But even at this level the short distance behavior
is affected. This means that in a less approximated scheme something could
change the short distance cutoff of Eq.$\left(  \ref{e36}\right)  $.

\item[\textbf{Remark 2}] We have hitherto considered the wormhole practically
as a black hole with a horizon located at $r_{w}$. However if one deals with
traversable wormholes, the brick wall is softened in a logarithmic divergence,
due to the horizon absence.
\end{description}

\section{The brick wall model and the foam}

\label{p3}In previous section, we have reproduced the 't Hooft brick wall
result by fixing the background geometry of one wormhole which behaves as a
black hole. In this section, we consider the idea that the divergent horizon
entropy may be affected by the presence of nonperturbative structure in the
vacuum at the Planck length scales. Instead of using the Crane-Smolin virtual
black holes, we consider as a natural candidate for such structure, a
distribution of virtual wormholes suggested by the uncertainty principle. Thus
in the spacelike slice $\Sigma,$ we consider the distribution $\rho\left(
r_{w}\right)  dr_{w}$ with radii between $r_{w}$ and $r_{w}+dr_{w}$ expressed
by%
\begin{equation}
\rho\left(  r_{w}\right)  =\left\{
\begin{array}
[c]{cc}%
0 & r_{w}>r_{h}\\
C/\left(  64e\pi^{2}r_{w}^{4}\right)   & l_{p}\leq r_{w}\leq r_{h},
\end{array}
\right.  \label{p41}%
\end{equation}
where $C$ is a dimensionless constant and where the Crane-Smolin distribution
$\left(  \ref{eqn:density}\right)  $ has been restricted only to the value
$q=0$ of the exponent corresponding to a scale invariant distribution. The
choice of the form of $\rho\left(  r_{w}\right)  $ is also suggested by the
behavior of the energy density of spacetime foam described by a collection of
non-interacting wormholes\cite{RIJMPD}\footnote{Actually the correct
expression found in Ref.\cite{RIJMPD} is%
\begin{equation}
\rho\left(  E_{w}\right)  \sim-\frac{\Lambda^{4}}{64e\pi^{2}}.
\end{equation}
The minus sign appears because the computation has been done looking at flat
space as the reference space.}%
\begin{equation}
\rho\left(  E_{w}\right)  \sim\frac{\Lambda^{4}}{64e\pi^{2}}%
\end{equation}
with the value of the cut-off $\Lambda$ substituted by $1/r_{w}$. To see if
spacetime foam modeled by wormholes affects the brick wall, we consider a
fixed black hole horizon $r_{h}$ filled by wormholes distributed following
Eq.$\left(  \ref{p41}\right)  $. Actually, in Eq.$\left(  \ref{p41}\right)  $,
we have considered a distribution of wormholes concentrated in a region of the
space inside the black hole horizon. This is suggested as a first
approximation by the results obtained in quantizing the entropy via
Bekenstein-Hawking relation of Eq.$\left(  \ref{eqn:S_BH}\right)
$\cite{RIJMPA}. However, nothing prevents to consider the whole space $\Sigma$
subjected to the distribution $\rho\left(  r_{w}\right)  $. We recall that our
main purpose is to see if and how the brick wall divergence is modified by an
underlying nontrivial spacetime structure. Since we have assumed that
$\Lambda\left(  r_{w}\right)  $ is a constant, without loss of generality and
for future purposes we can modify Eq.$\left(  \ref{p41}\right)  $ in the
following way%
\begin{equation}
\rho\left(  r_{w}\right)  =\left\{
\begin{array}
[c]{cc}%
0 & r_{w}>r_{h}\\
C^{\prime}\exp\left(  \Lambda\left(  r_{w}\right)  \right)  /\left(
64e\pi^{2}r_{w}^{4}\right)   & l_{p}\leq r_{w}\leq r_{h},
\end{array}
\right.
\end{equation}
We proceed to compute the total black hole entropy beginning with the
expression $\left(  \ref{e33a}\right)  $ obtained in the previous section%
\begin{equation}
S\left(  h,r_{h},r_{w}\right)  =\frac{16\pi^{3}}{90}\frac{g\left(
h,r_{h},r_{w}\right)  }{4\beta_{\infty,r_{w}}^{3}\kappa_{w}^{2}},\label{p42}%
\end{equation}
where the integration range is now $l_{p}\leq r_{w}\leq r_{h}$ and%
\begin{equation}
g\left(  h,r_{h},r_{w}\right)  =\int_{r_{h}}^{r_{h}+h}\frac{r^{2}}{\left(
r-r_{w}\right)  ^{2}}dr.\label{p42a}%
\end{equation}
Eq.$\left(  \ref{p42}\right)  $ describes the entropy generated by one
wormhole with a throat located at $r_{w}$ with respect to the black hole
horizon $r_{h}$. The main difference between Eq.$\left(  \ref{p42a}\right)  $
and Eq.$\left(  \ref{e34}\right)  $ is in the integration limits: indeed in
Eq.$\left(  \ref{e34}\right)  $, we have considered the wormhole exactly like
a black hole, while in Eq.$\left(  \ref{p42a}\right)  $, the black hole is
formed by wormholes of smaller radius. It is obvious that the typical brick
wall divergence in Eq.$\left(  \ref{p42a}\right)  $ is absent. However, this
is not the complete and correct expression of the brick wall calculation,
because we have to sum over all wormholes contributing the black hole. If we
define $N_{w}\equiv N_{w}\left(  A_{r_{w}}\right)  $ as the number of
wormholes filling the area of a two-sphere $S^{2}$ of radius $r_{w}$, then the
variation of black hole entropy due to a variation in the number of wormholes
filling the black hole area is%
\begin{equation}
dS\left(  r_{h}\right)  =S\left(  h,r_{h},r_{w}\right)  dN_{w},
\end{equation}
we can write%
\begin{equation}
dS\left(  r_{h}\right)  =S\left(  h,r_{h},r_{w}\right)  \frac{dN_{w}%
}{dA_{r_{w}}}dA_{r_{w}}%
\end{equation}
and the total entropy is%
\begin{equation}
S\left(  r_{h}\right)  =\int_{A_{l_{p}}}^{A_{r_{h}}}S\left(  h,r_{h}%
,r_{w}\right)  \frac{dN_{w}}{dA_{r_{w}}}dA_{r_{w}}.\label{p43}%
\end{equation}
If we assume that for every $r_{w}$%
\begin{equation}
T_{\infty,r_{w}}=\frac{\kappa_{w}}{2\pi},
\end{equation}
then exchanging the integration order, Eq.$\left(  \ref{p43}\right)  $ becomes%
\begin{equation}
\int_{A_{l_{p}}}^{A_{r_{h}}}S\left(  h,r_{h},r_{w}\right)  \frac{dN_{w}%
}{dA_{r_{w}}}dA_{r_{w}}=\frac{\pi}{90}\int_{r_{h}}^{r_{h}+h}r^{2}\left[
\int_{A_{l_{p}}}^{A_{r_{h}}}\frac{\kappa_{w}}{\left(  r-r_{w}\right)  ^{2}%
}\frac{dN_{w}}{dA_{r_{w}}}dA_{r_{w}}\right]  dr.\label{p44}%
\end{equation}
The last point concerns the number of wormholes per unit area $dN_{w}%
/dA_{r_{w}},$ representing the maximum number of wormholes that can be stored
in an area of radius $r_{w}$ beginning with an area of Planckian size
$A_{l_{p}}$. This number can be computed by means of the distribution
$\rho\left(  r_{w}\right)  $. This is simply obtained by the following
expression%
\begin{equation}
\frac{dN_{w}}{dA_{r_{w}}}=\int_{A_{l_{p}}}^{A_{r_{w}}}\rho\left(
r_{w}^{\prime}\right)  dA_{r_{w}^{\prime}},\label{p44a}%
\end{equation}
which affects the brick wall behavior. Indeed the expression in
Eq.$\left( \ref{p44a}\right)  $ establishes the counting of the
constituents of the horizon with respect to the area.
Unfortunately, a direct comparison with the usual expressions
obtained in Refs.\cite{Xiang Li,RenQinChun,Rama} is not immediate,
because we have worked in terms of radial coordinates and not with
the momentum representation. A comparison should be possible if
Eq.$\left( \ref{p44a}\right)  $ should be Fourier transformed in
terms of the momentum $p$. However, this goes beyond the purpose
of this paper. Thus Eq.$\left(
\ref{p44}\right)  $ can be written as%
\begin{equation}
C^{\prime}\left[  1-b^{\prime}\left(  r_{w}\right)  \right]  \frac{\pi}%
{90}\int_{r_{h}}^{r_{h}+h}r^{2}\left[  \frac{\left(  4\pi\right)  ^{2}}%
{64e\pi^{2}}\int_{l_{p}}^{r_{h}}\frac{dr_{w}}{\left(  r-r_{w}\right)  ^{2}%
}\left[  \frac{1}{l_{p}^{2}}-\frac{1}{r_{w}^{2}}\right]  \right]
dr,\label{p45}%
\end{equation}
where we have used the explicit expression of $\kappa_{w}$ given by
Eq.$\left(  \ref{e32}\right)  $. In Eq.$\left(  \ref{p45}\right)  $, we have
set%
\begin{equation}
\frac{C^{\prime}}{e}\left[  1-b^{\prime}\left(  r_{w}\right)  \right]  =1.
\end{equation}
This is true for Schwarzschild-like wormholes where $b^{\prime}\left(
r_{w}\right)  $ vanishes or for wormholes with $b^{\prime}\left(
r_{w}\right)  =const<1$. When $b^{\prime}\left(  r_{w}\right)  $ is a function
of $r_{w}$, the result depends on a case to case. To leading order in $h$, one
gets%
\begin{equation}
S\left(  r_{h}\right)  =\frac{\pi}{360}\left[  \frac{r_{h}^{2}}{l_{p}^{2}%
}-1\right]  \ln\left(  \frac{1}{h}\right)  +\text{ finite terms as
}h\rightarrow0.
\end{equation}
If $r_{h}^{2}/l_{p}^{2}\gg1$ then%
\begin{equation}
S\left(  r_{h}\right)  =\frac{A}{1440l_{p}^{2}}\ln\left(  \frac{1}{h}\right)
.\label{p46}%
\end{equation}

\section{Summary and discussion}

\label{p4}In this paper, we have examined the possibility that a complicated
structure like a foamy spacetime may affect the ultra-violet behavior of the
brick wall. What we have obtained is a softening of the divergence that is
turned from a linear to a logarithmic type. Although a certain number of
assumptions has been considered in using the wormhole metric $\left(
\ref{e31}\right)  $, the result seems to be quite general. Indeed, the model
of spacetime foam picture we have used depends strictly on the constituents,
which in our case, are Schwarzschild-like wormholes. However, we have found
that Schwarzschild-Anti-de Sitter wormholes could be used as representatives
of the foam. If this choice is adopted, in that case the brick wall will
exhibit a completely different behavior due to the different form of
$b^{\prime}\left(  r_{w}\right)  $ involved. Thus, we expect that Eq.$\left(
\ref{p46}\right)  $ can be valid also for more complicated black holes.

\end{document}